\documentclass{jfm}

\usepackage{graphicx}
\usepackage{natbib}
\usepackage{color}
\usepackage{lineno}
\usepackage{pdfpages}

\begin{document}

\newtheorem{lemma}{Lemma}
\newtheorem{corollary}{Corollary}

\shorttitle{Squirmers with swirl -- {\it Volvox} swimming} 
\shortauthor{T.~J.~Pedley et al.} 

\title{Squirmers with swirl -- a model for \emph{Volvox} swimming}

\author
 {
 T.~J.~Pedley\aff{1}
  \corresp{\email{t.j.pedley@damtp.cam.ac.uk}},
  D.~R.~Brumley\aff{2,3}
  \and 
  R.~E.~Goldstein\aff{1}
  }

\affiliation
{
\aff{1}
Department of Applied Mathematics and Theoretical Physics, University of Cambridge, Centre for Mathematical Sciences, 
Wilberforce Road, Cambridge CB3 0WA, UK
\aff{2}
Ralph M. Parsons Laboratory, Department of Civil and Environmental Engineering, Massachusetts Institute of Technology, 
Cambridge, MA 02139, USA
\aff{3}
Department of Civil, Environmental and Geomatic Engineering, ETH Zurich, 8093 Zurich, Switzerland
}

\maketitle

\begin{abstract}
Colonies of the green alga \emph{Volvox} are spheres that swim through the beating of pairs of flagella on their surface somatic cells. 
The somatic cells themselves are mounted 
rigidly in a polymeric extracellular matrix, fixing the orientation of the flagella so 
that they beat approximately in a meridional plane, with axis of symmetry in the swimming direction, but with a roughly 15 degree 
azimuthal offset which results in the eponymous rotation of the colonies about a body-fixed axis. Experiments on colonies held stationary 
on a micropipette show that the beating pattern takes the form of a symplectic metachronal wave (\cite{Brumley:2012}). Here we 
extend the Lighthill/Blake axisymmetric, Stokes-flow model of a free-swimming spherical squirmer  (\cite{Lighthill:1952,Blake:squirmer}) 
to include azimuthal swirl. The measured kinematics of the metachronal wave for 60 different colonies are used to calculate the 
coefficients in the eigenfunction expansions and hence predict the mean swimming speeds and rotation rates, proportional to the 
square of the beating amplitude, as functions of colony radius. As a test of the squirmer model, the results are compared with 
measurements (\cite{Drescher:2009vn}) of the mean swimming speeds and angular velocities of a different set of 220 colonies, 
also given as functions of colony radius. The predicted variation with radius is qualitatively correct, but the model underestimates 
both the mean swimming speed and the mean angular velocity unless the amplitude of the flagellar beat is taken to be larger than 
previously thought. The reasons for this discrepancy are discussed.

\end{abstract}


\section{Introduction}\label{sec:1}

\begin{figure}
\begin{center}
\includegraphics[width=0.3\textwidth]{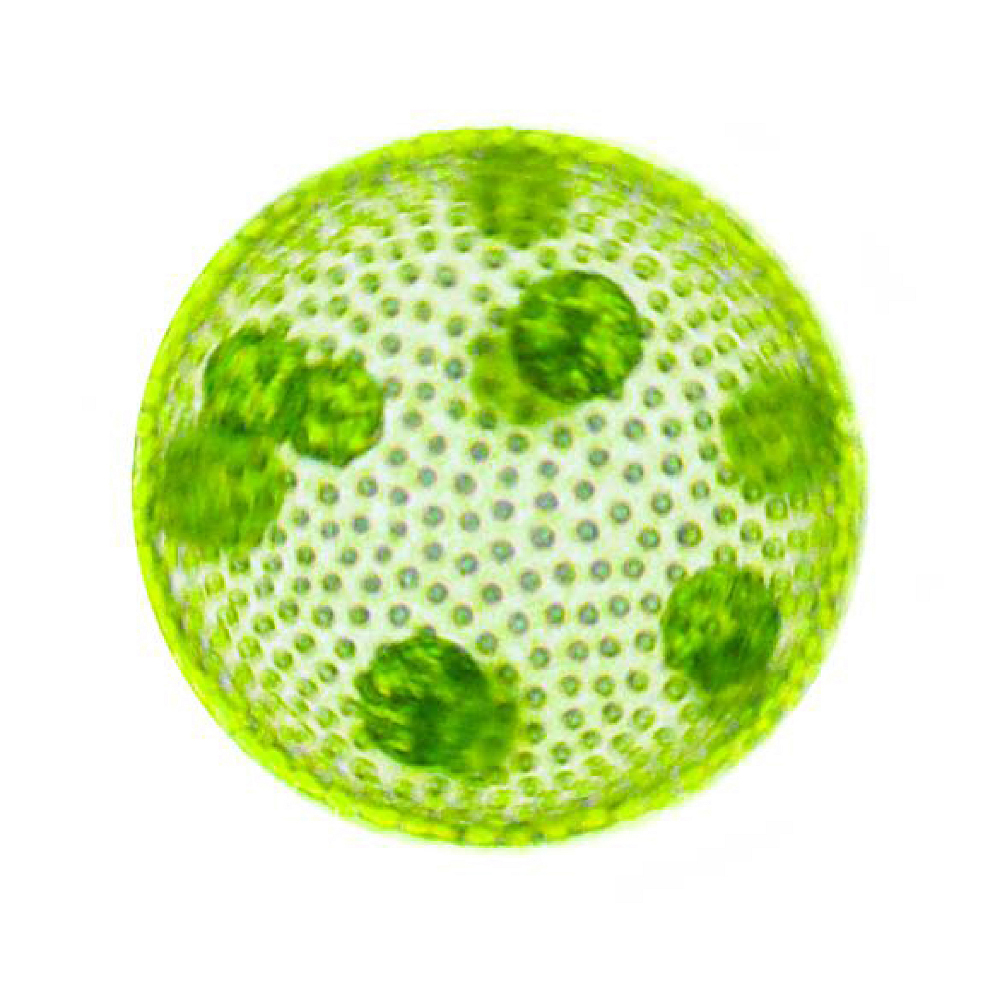}
\end{center}
\caption{A colony of {\it Volvox carteri}. Small green dots are the somatic cells on the outside ($2,000-6,000$ for {\it V. carteri}); larger 
green spheroids are the interior daughter colonies. The photograph is taken from above, as the colony swims upwards towards the camera.} \label{volvox_image}
\end{figure}

\emph{Volvox} is a genus of algae with spherical, free-swimming colonies consisting of up to 50,000 surface somatic cells embedded in an 
extracellular matrix and a small number of interior germ cells which develop to become the next generation (figure~\ref{volvox_image}). 
Discovered by \citet{vanLeeuwenhoek:1700}, who marveled at their graceful swimming, it
was named by \citet{Linnaeus:1758} for its characteristic spinning motion about
the anterior-posterior axis.  Each somatic cell has two flagella that all beat more or less in planes that are offset from purely meridional 
planes by an angle of $10^{\circ} - 20^{\circ}$; it is believed that this offset causes the observed rotation.
The power stroke of a flagellum's beat is directed towards the rear - i.e. from the `north pole' towards the `south pole', apart from the 
angular offset.  The colonies are about 0.3\% denser than water, and swim upwards in still water, parallel to the axis of symmetry, because 
the relatively dense interior cells are clustered towards the posterior; when the axis is deflected from vertical the colony experiences a 
restoring gravitational torque that competes with a viscous torque to right the colony on a timescale
of $\sim 10$ s.  

\subsection{Experimental background}

\begin{figure}
\begin{center}
\includegraphics[width=0.8\textwidth]{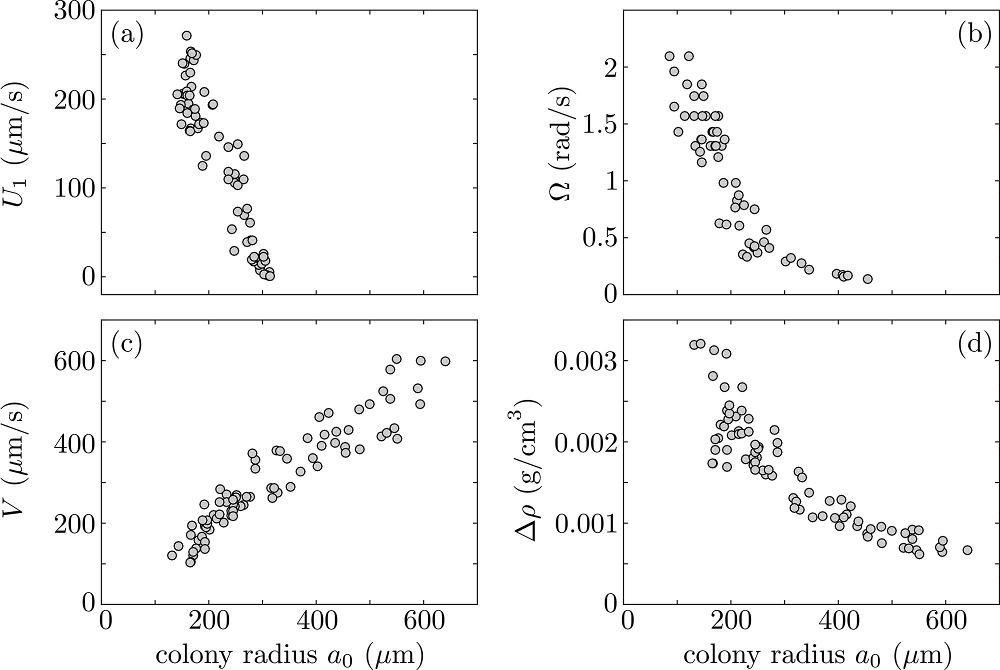}
\end{center}
\caption{Swimming properties of {\it V. carteri} as a function of colony radius $a_0$. Measured values of the (a) upswimming 
speed $U_1$, (b) angular velocity $\Omega$, and (c) sedimentation speed $V$, as well as (d) the deduced density offset 
$\Delta \rho = 9 \mu V / 2 g a_0^2$ compared to the surrounding medium. Adapted from \cite{Drescher:2009vn}.} \label{Drescher_results}
\end{figure}

During its 48-hour life cycle, the size of a \emph{Volvox} colony increases, though the number and size of somatic cells do not. Thus one 
would expect the sedimentation speed $V$ of a colony whose swimming was arrested to increase with colony radius $a_0$, while its 
upswimming speed $U_1$ would decrease, both because of the increase in $V$ and because, even if it were neutrally buoyant, one 
would expect the viscous drag to increase with size and hence the swimming speed $U$ to decrease. Presumably the angular velocity 
about the axis, $\Omega$, would also decrease. Drescher {\it et al.} measured the swimming speeds, sedimentation speeds, and 
angular velocities of 78, 81 and 61 colonies of \emph{Volvox carteri} respectively, ranging in radius from about 100~$\mu$m to about 
500~$\mu$m. The results are shown in figure~\ref{Drescher_results}, where indeed both $U_1$ and $\Omega$ are seen to decrease 
with $a_0$, while $V$ increases. The expected swimming speed if the colony were neutrally buoyant would be $U = U_1 + V$ (linearity 
is expected because the Reynolds number of even the largest colony is less than 0.1, so the fluid dynamics will be governed by the Stokes 
equations).

\begin{figure}
\begin{center}
\includegraphics[width=0.35\textwidth]{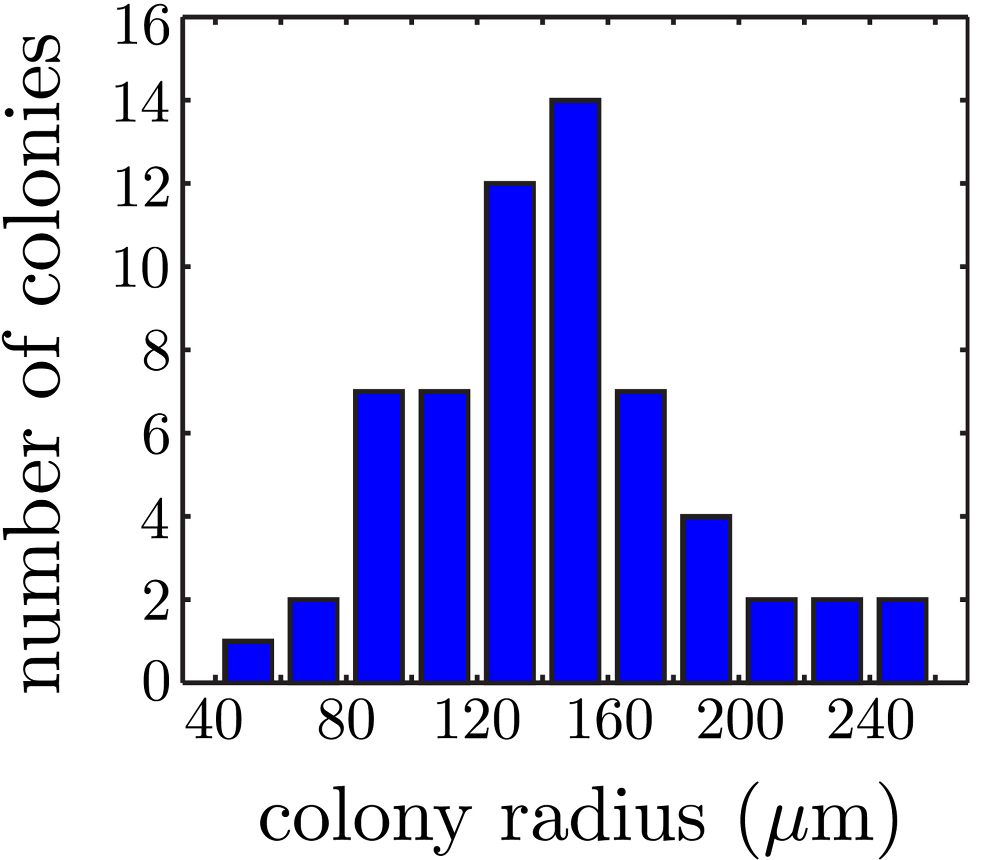}
\end{center}
\caption{Distribution of colonies by radius, for which the metachronal wave properties are characterized. Adapted from figure~1(b) of  \cite{Brumley:2015MW}.} \label{radius_pdf}
\end{figure}

The purpose of this paper is to describe a model for \emph{Volvox} swimming from which both $U$ and $\Omega$ can be predicted, and 
to compare the predictions with the experiments of figure~\ref{Drescher_results}. The input to the model will be the fluid velocities generated 
by the flagellar beating as measured by \cite{Brumley:2012,Brumley:2015MW}. Detailed measurements were made of the time-dependent 
flow fields produced by the beating flagella of numerous \emph{V. carteri} colonies. Individual colonies were held in place on a micro-pipette 
in a $25\times25\times5$~mm glass observation chamber; the colonies were attached at the equator and arranged so that the symmetry 
axis of a colony was perpendicular both to the pipette and to the field of view of the observing microscope. The projection of the flow field 
onto the focal plane of the microscope was visualised by seeding the fluid medium with $0.5 \mu$m polystyrene microspheres at a volume 
fraction of $2\times 10^{-4}$, and thirty-second-long high speed movies were taken. The (projected) velocity field was measured using 
particle image velocimetry (PIV); a total of 60 different colonies were investigated, ranging in radius from $48~\mu$m to $251~\mu$m 
(mean $144 \pm 43~\mu$m), the distribution of which is shown in figure~\ref{radius_pdf}. 

\begin{figure}
\begin{center}
\includegraphics[width=\textwidth]{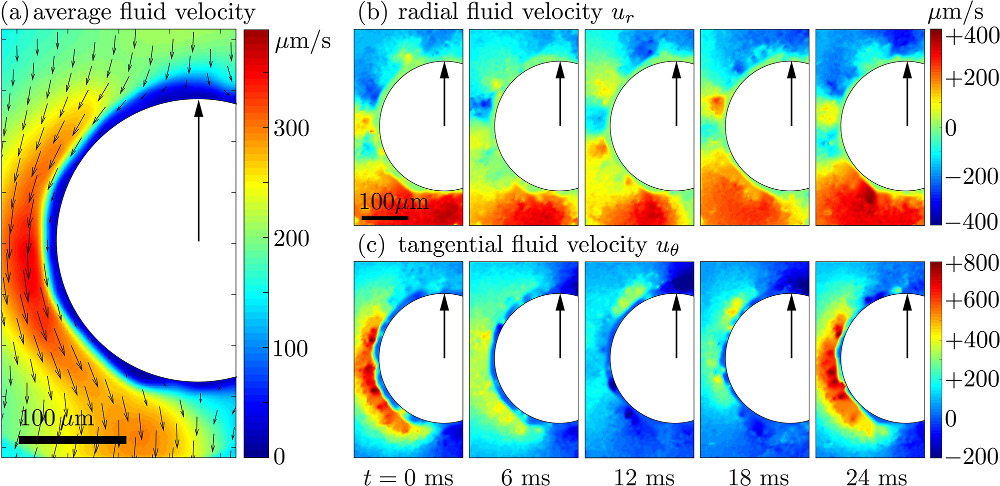}
\end{center}
\caption{Experimental flow fields. (a) Magnitude (colour) and direction (arrows) of the time-averaged velocity field measured with PIV. 
Radial (b) and tangential (c) components of the fluid velocity field shown at various times through one flagellar beating cycle. Part a is 
adapted from figure~1(c) of \cite{Brumley:2015MW}.} \label{exp_flow}
\end{figure}

One example of the time-averaged magnitude of the velocity distribution is shown in figure~\ref{exp_flow}(a). This is a maximum near the 
equator because the flagellar beating drives a non-zero mean flow past the colony, parallel to the axis of symmetry and directed from front 
to back. This is consistent with the fact that untethered colonies swim forwards, parallel to the axis.

\begin{figure}
\begin{center}
\includegraphics[width=0.9\textwidth]{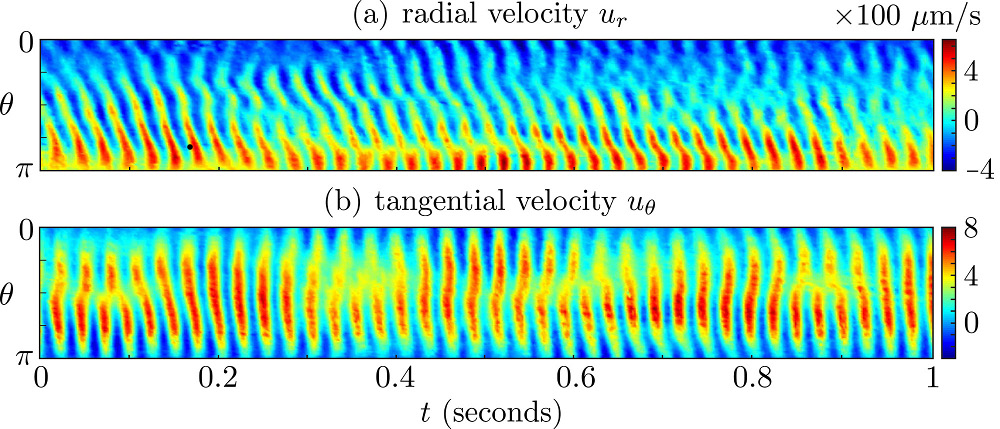}
\end{center}
\caption{Kymographs of radial (a) and tangential (b) velocity around {\it Volvox} colonies, measured at a radius of $r = 1.3 \times a_0$. 
Adapted from figure~2 of \cite{Brumley:2015MW}.} \label{exp_kymographs}
\end{figure}

More interesting are the perturbations to this mean flow. Time-dependent details of velocity field can be seen in supp. mat. movies S1 
and S2. Close to the colony surface, backwards and forwards motion, driven by the beating flagella, can be clearly seen; further away 
the flow is more nearly steady. Figure~\ref{exp_flow} contains a series of snapshots showing unsteady components of the (b) radial 
velocity, $u_{r}$, and (c) tangential velocity, $u_\theta$. It is immediately evident that the maximum of radial velocity propagates as a 
wave from front to back, in the same direction as the power stroke of the flagellar beat -- a symplectic metachronal wave 
(\cite{Sleigh:1960ys}). This is further demonstrated in figure~\ref{exp_kymographs} which shows kymographs of $u_{r}$ and  
$u_\theta$ measured at a distance $r = 1.3 \times a_0$ from the colony surface: the propagating wave is clearly seen in 
figure~\ref{exp_kymographs}(a), while figure~\ref{exp_kymographs}(b) suggests that the tangential velocity behaves more 
as a standing wave, dominated by the power stroke near the equator. 

The results of \cite{Brumley:2012} show that a good fit to to the observations of the radial velocity perturbations is given by the 
following simple form:
\begin{equation}\label{eq:1.1}
u'_r |_{r=1.3a_0} = \sigma a_0 \epsilon \cos{(k\theta_0 - \sigma t)},
\end{equation}
where $\theta_0$ is the polar angle, $k,\sigma$ are the wave-number and frequency of the wave, and $\epsilon$ is an amplitude 
parameter. The mean values of $k, \sigma, \epsilon$ over all the colonies observed were $k = 4.7$, $\sigma = 203~$rad~s$^{-1}$, 
$\epsilon \approx 0.035  $. Such data for each colony measured will make up the full input to our model below.

\subsection{Theoretical background}

\begin{figure}
\begin{center}
\includegraphics[height=7.5cm]{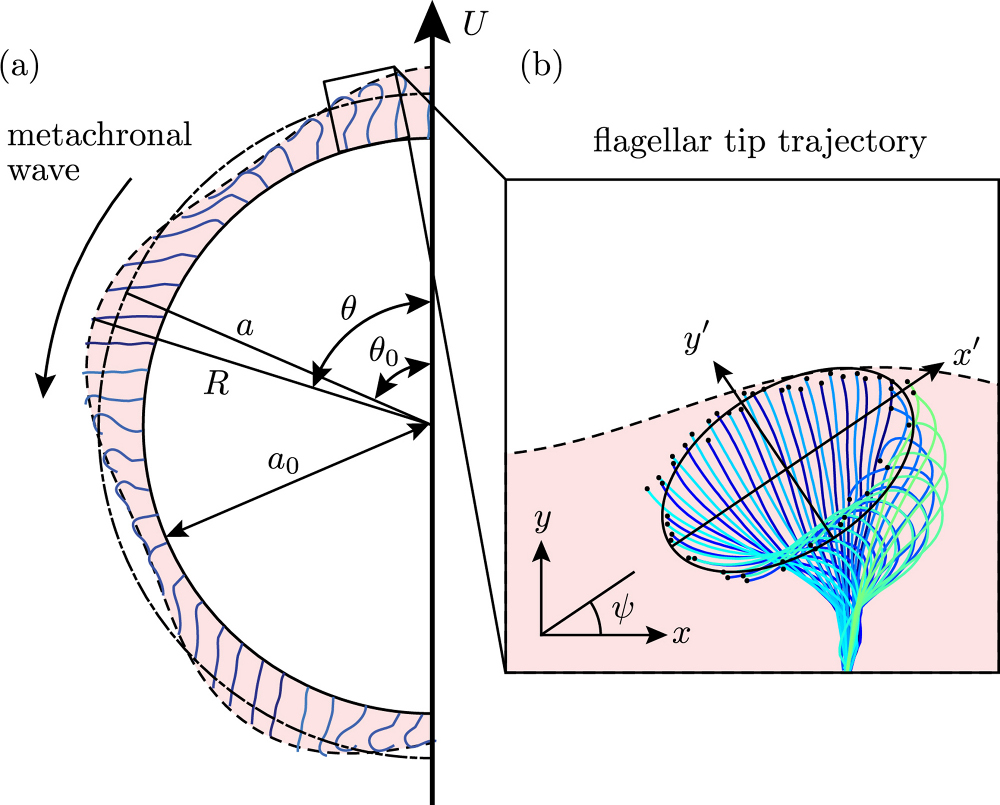}
\end{center}
\caption{(a) Schematic diagram of a spherical {\it Volvox} colony at one instant in time, with beating flagella and the envelope of flagellar tips. 
The radius of the extracellular matrix in which the flagella are embedded is $a_0$. The mean radius of the envelope is $a$; $(R,\theta)$ are 
the coordinates of a surface element whose average position is $(a,\theta_0)$ [Adapted from \cite{Blake:squirmer}, but replotted with the 
experimentally-determined metachronal wavenumber]. (b) Measured tip trajectory over multiple beats of a single{\it Volvox} flagellum. The 
trajectory is fitted with an ellipse, which is rotated at an angle $\psi$ with respect to the local colony surface.} 
\label{envelope}
\end{figure}

The model will be an extension to the swirling case of the spherical envelope (or `squirmer') model for the propulsion of ciliated protozoa 
introduced by \cite{Lighthill:1952} and \cite{Blake:squirmer}. When the surface of a cell is densely covered with beating cilia, as for the 
protist \emph{Opalina} for example, it is a very good approximation to treat the flow around it as being driven by the displacement of a 
stretching sheet, attached to the tips of all the cilia and moving with them. The sheet will undergo radial and tangential wave-like displacements, 
and it needs to stretch to accommodate temporal variations between the displacements of neighbouring cilia tips (figure~\ref{envelope}(a)). 
In the case of \emph{Volvox} the tips of the beating flagella are not very close together; for a colony of radius $200~\mu$m, the average 
spacing between somatic cells is $\sim 20~\mu$m, comparable with the flagellar length, $\langle L \rangle =  19.9~ \mu$m (\cite{Brumley2014}), 
so the envelope model may well be somewhat inaccurate. As indicated above, the new feature of our model is the introduction of azimuthal swirl to the envelope model.

The theory will be given in the next two sections, first extending the Lighthill-Blake model to include swirl, and second applying the model to 
\emph{Volvox} on the basis of the data of \cite{Brumley:2012}. The objective is to calculate the mean swimming speed $\bar{U}$ and mean 
angular velocity $\bar{\Omega}$, and test the model by comparison with the measurements of \cite{Drescher:2009vn}. The final section will 
include a discussion of discrepancies and the model's limitations.

\section{Theory for squirmers with swirl}\label{sec:2}

In the original, zero-Reynolds-number, spherical-envelope model of ciliated micro-organisms (\cite{Lighthill:1952,Blake:squirmer}), the radial 
and tangential Eulerian velocity components $(u_r, u_\theta)$ are written as infinite series of eigensolutions of the Stokes equation:
\begin{subequations}\label{eq:2.1}
\begin{eqnarray}\label{eq:2.1a}
u_r(r,\theta_0) &&= -U\cos{\theta_0} + A_0 \frac{a^2}{r^2} P_0 + \frac{2}{3}(A_1 +  B_1) \frac{a^3}{r^3} P_1 +  \\ 
&& \sum_{n=2}^{\infty}\left[\left(\frac{1}{2} n \frac{a^n}{r^n} - (\frac{1}{2}n - 1)\frac{a^{n+2}}{r^{n+2}}\right)A_n P_n  + 
\left(\frac{a^{n+2}}{r^{n+2}} - \frac{a^n}{r^n}\right)B_nP_n\right] \nonumber
\end{eqnarray}
\begin{eqnarray}\label{eq:2.1b}
u_\theta (r,\theta_0) &&= U\sin{\theta_0} + \frac{1}{3}(A_1 + B_1)\frac{a^3}{r^3}V_1 +  \\ 
&&\sum_{n=2}^{\infty}\left[\left(\frac{1}{2} n \frac{a^{n+2}}{r^{n+2}} - (\frac{1}{2}n - 1)\frac{a^n}{r^n}\right)B_n V_n  
+ \frac{1}{2} n(\frac{1}{2} n - 1) \left(\frac{a^n}{r^n} - \frac{a^{n+2}}{r^{n+2}}\right)A_nV_n\right], \nonumber
\end{eqnarray}
\end{subequations}
assuming axial symmetry. Here $(r,\theta_0)$ are spherical polar co-ordinates, the $P_n(\cos{\theta_0})$ are Legendre 
polynomials, and
\begin{equation}\label{eq:2.2}
V_n (\cos{\theta_0}) = \frac{2}{n(n+1)} \sin{\theta_0} P_n'(\cos{\theta_0}).
\end{equation}
A trace of a typical flagellar beat is shown in figure~\ref{envelope}(b), adapted from \cite{Brumley2014}, where it can be seen 
that the trajectory of the tip is approximately elliptical, with centre about two-thirds of the flagellar length from the surface of the 
extracellular medium. Thus $a$ is taken to be the mean radius of a flagellar tip, so we take $a\approx a_0 + 2L/3$, where $L$ is 
the length of a flagellum. With the origin fixed at the centre of the sphere, $-U(t)$ is the speed of the flow at infinity (i.e. $U$ is the 
instantaneous swimming speed of the sphere). If the sphere is taken to be neutrally buoyant, it experiences no external force, so the 
Stokeslet term must be zero, and
\begin{equation}\label{eq:2.3}
U = \frac{2}{3}B_1 - \frac{1}{3}A_1
\end{equation}
(\cite{Blake:squirmer}). Corresponding to the velocity field (\ref{eq:2.1}), the velocity components on the sphere $r=a$ are
\begin{equation}\label{eq:2.4}
u_r (a,\theta_0) = \sum_{n=0}^{\infty}A_n (t)P_n (\cos{\theta_0}), ~~ u_\theta (a,\theta_0) =\sum_{n=1}^{\infty}B_n 
(t)V_n (\cos{\theta_0}) .
\end{equation}
From this we can see that $A_1$ should be zero, because it corresponds to longitudinal translation of the centre, which is 
incorporated into $U$. However, we follow Lighthill (1952) and not \cite{Blake:squirmer} in retaining a non-zero $A_0$. Blake 
wished to prohibit any volume change in his squirmers, but we note that if, say, all the flagella beat synchronously, the envelope 
of their tips would experience a small variation in volume, so $A_0$ should not be zero.

The surface velocities in Eq.~(\ref{eq:2.4}) must in fact be generated by the motion of material elements of the spherical envelope, 
representing the tips of the beating flagella. In the Lighthill-Blake analysis, the envelope is represented by the following expressions 
for the Lagrangian co-ordinates $(R,\theta)$ of the material elements:
\begin{subequations}\label{eq:2.5}
\begin{equation}\label{eq:2.5a}
R - a = a\epsilon\sum_{n=0}^{\infty} \alpha_n (t) P_n (\cos{\theta_0})
\end{equation}
\begin{equation}\label{eq:2.5b}
\theta - \theta_0 = \epsilon\sum_{n=1}^{\infty} \beta_n (t) V_n (\cos{\theta_0}).
\end{equation}
\end{subequations}
The functions $\alpha_n(t)$ and $\beta_n(t)$ are supposed to be oscillatory functions of time with zero mean, and the amplitude of the 
oscillations, $\epsilon$, is taken to be small. The most intricate part of the theory is the calculation of the $A_n$ and $B_n$ in Eq.~(\ref{eq:2.4}) 
in terms of the $\alpha_n$ and $\beta_n$ in Eq.~(\ref{eq:2.5}). This will be outlined below.

The new feature that we introduce in this paper is to add axisymmetric swirl velocities and azimuthal ($\phi$) displacements to the above. 
The $\phi$-component of the Stokes equation is
\begin{equation}\label{eq:2.6}
\nabla ^2 u_\phi - \frac{u_\phi}{r^2 \sin^2{\theta_0}} = 0
\end{equation}
and the general solution that tends to zero at infinity is
\begin{equation}\label{eq:2.7}
u_\phi (r,\theta_0) = \sum_{n=1}^{\infty} a \ C_n \frac{a^{n+1}}{r^{n+1}} V_n (\cos{\theta_0}),
\end{equation}
equal to
\begin{equation}\label{eq:2.8}
u_\phi (a,\theta_0) = \sum_{n=1}^{\infty} a \ C_n V_n (\cos{\theta_0})
\end{equation}
on $r=a$. This solution was also given by \citet{PakLauga2014}, who also considered non-axisymmetric squirming and swirling, and calculated
the translational and angular velocities corresponding to any distribution of velocities on $r=a$.  
Now the total torque about the axis of symmetry is $-8\pi \mu a^3 C_1$ and, since the sphere is our model for a 
free-swimming \emph{Volvox} colony, this, like the total force, must be zero - i.e.
\begin{equation}\label{eq:2.9}
C_1 \equiv 0.
\end{equation}
Analogous to Eq.~(\ref{eq:2.5}), the $\phi$-displacement of the material point $(R,\theta,\phi)$ on the spherical envelope is taken 
to be $\phi - \phi_0$ where
\begin{equation}\label{eq:2.10}
(\phi - \phi_0)\sin{\theta_0} = \int {\Omega dt} \sin{\theta_0} + \epsilon \sum_{n=1}^{\infty} \gamma_n (t) V_n(\cos{\theta_0}).
\end{equation}
Here $\phi _0$ is fixed on the rotating sphere, and $\Omega$ is the instantaneous angular velocity of the sphere.

The relations between the Eulerian velocities (\ref{eq:2.1}), (\ref{eq:2.7}) and the Lagrangian displacements (\ref{eq:2.5}), (\ref{eq:2.10}), 
from which $A_n$, $B_n$, $C_n$ and $U$, $\Omega$ are to be derived from $\alpha_n, \beta_n, \gamma_n$, are:
\begin{equation}\label{eq:2.11}
u_r(R,\theta) = \dot{R}, ~~ u_\theta (R,\theta) = R\dot{\theta}, ~~ u_\phi (R,\theta) = R\sin{\theta}\dot{\phi},
 \end{equation}
where an overdot represents the time derivative. \cite{Blake:squirmer} performed the analysis for the $r$- and $\theta$-velocities; here 
we illustrate the method by deriving the relation between the $C_n$ and the $\gamma_n$.

The analysis is developed in powers of the amplitude $\epsilon$, so we take
\begin{subequations}
\begin{equation}\label{eq:2.12a}
C_n = \epsilon C_n^{(1)} + \epsilon^2 C_n^{(2)} + ...
\end{equation}
\begin{equation}\label{eq:2.12b}
\Omega = \epsilon \Omega^{(1)} + \epsilon^2 \Omega^{(2)} + ... .
\end{equation}
\end{subequations}
At leading order, $O(\epsilon)$, (\ref{eq:2.11}c) and (\ref{eq:2.10}) give
\begin{equation}\label{eq:2.13}
C_1^{(1)} = \Omega^{(1)} + \dot{\gamma}_1 , ~~ C_n^{(1)} = \dot{\gamma}_n ~~(n>1).
\end{equation}
Immediately, therefore, we see from (\ref{eq:2.9}) that $\Omega^{(1)} = -\dot{\gamma}_1$, which has zero mean, so the mean angular 
velocity, like the mean translational speed, is $O(\epsilon^2)$. At second order, the fact that $(R,\theta) \neq (a,\theta_0)$ is important in the 
expression for the velocity field:
\begin{equation}\label{eq:2.14}
u_\phi (R,\theta) = u_\phi (a,\theta_0) + (R - a) \frac{\partial{u_\phi}}{\partial{r}}|_{a,\theta_0} + 
(\theta - \theta_0) \frac{\partial{u_\phi}}{\partial{\theta_0}}|_{a,\theta_0} + ... = R\sin{\theta}\dot{\phi}.
\end{equation}
Substituting for $R,\theta ,\phi$ gives:
\begin{eqnarray}\label{eq:2.15}
\sum_{n=1}^{\infty}(\epsilon C_n^{(1)} && + \epsilon^2 C_n^{(2)}) V_n - \\ && \epsilon^2 \sum_{n=0}^{\infty} \alpha_n P_n \sum_{m=2}^{\infty}(m+1)\dot{\gamma}_m V_m  + \epsilon^2 \sum _{n=1}^{\infty} \beta_n V_n \sum_{m=2}^{\infty}\dot{\gamma}_m \left(2P_m - \frac{\cos{\theta_0}}{\sin{\theta_0}} V_m\right) \nonumber \\ && = \epsilon \sin{\theta_0} \left(1 + \epsilon \sum_{n=0}^{\infty} \alpha_n P_n + \epsilon \frac{\cos{\theta_0}}{\sin{\theta_0}}\sum_{n=1}^{\infty} \beta_n V_n\right) \left(\Omega^{(1)} + \epsilon\Omega^{(2)} + \frac{1}{\sin{\theta_0}}\sum_{m=1}^{\infty} \dot{\gamma}_m V_m\right). \nonumber
\end{eqnarray}
Taking the $O(\epsilon^2)$ terms in this equation, multiplying by $\sin^2{\theta_0}$ and integrating from $\theta_0 = 0$ to $\theta_0 = \pi$ 
(recalling that $C_1^{(2)} = 0$), gives the following explicit expression for $\Omega^{(2)}$:
\begin{eqnarray}\label{eq:2.16}
\Omega^{(2)} && = -\frac{4}{5} \beta_1 \dot{\gamma}_2 + \sum_{n=2}^{\infty} \frac{3}{(2n+1)(2n+3)} [-(n+3) \alpha_n \dot{\gamma}_{n+1} + (n+2) \alpha_{n+1} \dot{\gamma}_n] \nonumber \\ && + \sum_{n=2}^{\infty} \frac{6}{(2n+1)(2n+3)(n+1)} [-(n+3) \beta_n \dot{\gamma}_{n+1} + (n-1) \beta_{n+1} \dot{\gamma}_n].
\end{eqnarray}
(Some of the required integrals of products of $P_n$ and $V_m$ are given in appendix~\ref{appA}). The corresponding result for the 
second order term in the translational velocity is:
\begin{eqnarray}\label{eq:2.17}
U^{(2)}/a = && \frac{2}{3} \alpha_0 \dot{\beta}_1 - \frac{8}{15}\alpha_2 \dot{\beta}_1 - \frac{2}{5} \dot{\alpha}_2 \beta_1 \nonumber \\ && + \sum_{n=2}^{\infty} \frac {(2n+4)\alpha_n \dot{\beta}_{n+1} - 2n\dot{\alpha}_n \beta_{n+1} - (6n+4)\alpha_{n+1} \dot{\beta}_n -(2n+4)\dot{\alpha}_{n+1}\beta_n}{(2n+1)(2n+3)} \nonumber \\ && + \sum_{n=1}^{\infty} \frac{4(n+2)\beta_n\dot{\beta}_{n+1} - 4n\dot{\beta}_n \beta_{n+1}}{(n+1)(2n+1)((2n+3)} \nonumber \\ && - \sum_{n=2}^{\infty} \frac{(n+1)^2 \alpha_n \dot{\alpha}_{n+1} - (n^2-4n-2)\alpha_{n+1} \dot{\alpha}_n}{(2n+1)(2n+3)}.
\end{eqnarray}
This is the formula given by \cite{Blake:squirmer}, except that he omitted the term involving $\alpha_0$ which Lighthill (1952) included; 
Lighthill omitted some of the other terms.

A shortcut to predicting $U$ and $\Omega$ was proposed by \cite{StoneSamuel}. They used the reciprocal theorem for Stokes flow to 
relate the translation and rotation speeds of a deformable body with non-zero surface velocity $\mathbf{u'}$ to the drag and torque on a 
rigid body of instantaneously identical shape, and derived the following results for a sphere of radius $a$, surface $S$:
\begin{subequations}
\begin{equation}\label{2.18a}
\mathbf{U}(t) = - \frac{1}{4\pi a^2} \int_S {\mathbf{u'} dS}
\end{equation}
\begin{equation}\label{2.18b}
\mathbf{\Omega}(t) = - \frac{3}{8\pi a^3} \int_S {\mathbf{n} \times \mathbf{u'} dS },
\end{equation}
\end{subequations}
where $\mathbf{n}$ is the outward normal to the sphere. From the first of these (\ref{eq:2.3}) follows. It turns out not to be so simple to 
use these results for squirmers with non-zero radial deformations, because of the need to calculate the drag to $O(\epsilon^2)$ for the 
rigid deformed sphere.

\section{Application to \emph{Volvox}}\label{sec:3}

In order to apply the above theory to \emph{Volvox}, we need to specify the $\alpha_n, \beta_n, \gamma_n$. This will be done by making 
use of the experimental results on the metachronal wave by Brumley et al (2012), which led to Eqn.~(\ref{eq:1.1}) for the radial velocity 
distribution on the envelope of flagellar tips, plus assumptions about the tangential and azimuthal displacements. Following Eq.~(\ref{eq:1.1}), 
we write the radial displacement as
\begin{equation}
R - a = a\epsilon \sin{(k\theta_0 - \sigma t)}, \label{eq:3.1}
\end{equation}
where $k$ is the wave number, $\sigma$ the radian frequency, and $\epsilon \ll 1$. Observations of flagellar beating show that a flagellar tip 
moves in an approximately elliptical orbit (see figure~\ref{envelope}(b)). Thus we may write
\begin{equation}
\theta - \theta_0 = \epsilon \delta \sin{(k\theta_0 - \sigma t -\chi)}, \label{eq:3.2}
\end{equation}
where figure~\ref{envelope}(b) suggests $\delta \approx 1.68$ and the phase difference $\chi \approx -\pi/2$. The observation that the 
plane of beating of the flagella is offset by $10^\circ - 20^\circ$ from the meridional plane suggests that the functional form of the 
$\phi$-displacement, 
relative to the rotating sphere, is also given by (\ref{eq:3.2}), multiplied by a constant, $\tau$, equal to the tangent of the offset angle.
Together, then, (\ref{eq:2.5}), (\ref{eq:2.10}), (\ref{eq:3.1}) and (\ref{eq:3.2}) give:
\begin{subequations}\label{eq:3.3}
\begin{equation}\label{eq:3.3a}
\alpha_0 (t) + \sum_{n=2}^{\infty} \alpha_n (t) P_n (\cos{\theta_0}) = \sin{(k\theta_0 - \sigma t)}
\end{equation}
\begin{equation}\label{eq:3.3b}
\sum_{n=1}^{\infty} \beta_n (t) V_n (\cos{\theta_0}) = \delta \sin{(k\theta_0 - \sigma t -\chi)}
\end{equation}
\begin{equation}\label{eq:3.3c}
\sum_{n=1}^{\infty} \gamma_n (t) V_n (\cos{\theta_0}) = \tau \delta \sin{(k\theta_0 - \sigma t -\chi)}.
\end{equation}
\end{subequations}

It can be seen immediately that $\gamma_n = \tau \beta_n$, so only (\ref{eq:3.3a}) and (\ref{eq:3.3b}) need to be solved for $\alpha_n$ 
and $\beta_n$. To do this requires expressions for $\sin{k\theta_0}$ and $\cos{k\theta_0}$ as series of both $P_n(\cos{\theta_0})$ and $V_n(\cos{\theta_0})$:
\begin{subequations}
\begin{equation}\label{eq:3.4a}
\sin{k\theta_0} = \sum_{n=0}^{\infty} a_n^{(s)} P_n (\cos{\theta_0}) = \sum_{n=1}^{\infty} b_n^{(s)} V_n (\cos{\theta_0})
\end{equation}
\begin{equation}\label{eq:3.4b}
\cos{k\theta_0} = \sum_{n=0}^{\infty} a_n^{(c)} P_n (\cos{\theta_0}) = \sum_{n=1}^{\infty} b_n^{(c)} V_n (\cos{\theta_0}).
\end{equation}
\end{subequations}
The results for $a_n^{(s)}$ etc (see appendix~\ref{appB}) are
\begin{subequations}\label{eq:3.5}
\begin{equation}\label{eq:3.5a}
a_n^{(s)}  = -k (2n+1) \big[ 1 + (-1)^{n+1}\cos k \pi \big]  \eta(k,n)
\end{equation}
\begin{equation}\label{eq:3.5b}
a_n^{(c)} = k (2n+1) (-1)^{n+1} \sin k \pi \ \eta(k,n)
\end{equation}
\begin{equation}\label{eq:3.5c}
b_n^{(s)} = \frac{1}{2} (-1)^{n+1} n(n+1)(2n+1) \sin k \pi \ \eta(k,n)
\end{equation}
\begin{equation}\label{eq:3.5d}
b_n^{(c)} = \frac{1}{2} n(n+1)(2n+1) \big[ 1+ (-1)^{n+1} \cos k \pi \big]  \eta(k,n)
\end{equation}
\end{subequations}
where
\begin{equation}\label{eq:3.6}
\eta(k,n) = \frac {\Gamma\left(\frac{n-k}{2}\right)\Gamma\left(\frac{n+k}{2}\right)}{16\Gamma\left(\frac{n+3-k}{2}\right)\Gamma\left(\frac{n+3+k}{2}\right)},
\end{equation}
and $k$ is assumed not to be an integer. It then follows from (\ref{eq:3.3}) that
\begin{subequations} \label{eq:3.7}
\begin{equation}\label{eq:3.7a}
\alpha_n(t) = k (-1)^{n+1} (2n+1) \big[ (-1)^n \cos \sigma t - \cos (\sigma t - k \pi) \big] \eta(k,n)
\end{equation}
\begin{equation}\label{eq:3.7b}
\beta_n(t) = \frac{\gamma_n}{\tau} = \frac{\delta}{2} (-1)^{n+1} n (n+1)(2n+1) \big[ (-1)^n \sin (\sigma t + \chi) - \sin (\sigma t + \chi - k\pi) \big] \eta(k,n).
\end{equation}
\end{subequations}

Now we can put Eqs.~(\ref{eq:3.7}) into Eqs.~(\ref{eq:2.16}) and (\ref{eq:2.17}), take the mean values, and obtain final results for the second order contributions to the mean angular and translational velocities:
\begin{eqnarray}\label{eq:3.8}
&& \bar{\Omega}^{(2)} = 36 \sigma  \tau \delta ^2 \eta(k,1)\eta(k,2) \sin k \pi \\ && + \frac{3}{2}\sigma\tau\delta \sin{k\pi}\sum_{n=2}^{\infty}\eta(k,n)\eta(k,n+1)(-1)^{n+1} (n+1)(n+2)[(2n+3)k\sin{\chi} + 2\delta n(n+1)], \nonumber
\end{eqnarray}
\begin{eqnarray}\label{eq:3.9}
&& \bar{U}^{(2)} = -2a\sigma \delta \eta(k,1)\eta(k,2) \sin{k\pi}(12\delta + \frac{9}{k}\sin{\chi})  + a\sigma\sin{k\pi}\sum_{n=2}^{\infty}(-1)^n \eta(k,n)\eta(k,n+1) \nonumber \\ && \times [2\delta^2 n(n+1)^2 (n+2) + 2k\delta (n+1)(2n^2 + 3n + 2)\sin{\chi} - k^2 (2n^2 - 2n -1)].
\end{eqnarray} 
We may note that calculations are made easier by recognising that
\begin{equation}\label{eq:3.10}
\eta(k,n)\eta(k,n+1) = \frac{1}{4((n+2)^2 - k^2)((n+1)^2 - k^2)(n^2 - k^2)}.
\end{equation}

We now put in parameter values obtained from the experiments of Brumley {\it et al}. (2012) and compare the predicted values of $\bar{U}$ and $\bar{\Omega}$ with the measurements of Drescher {\it et al}. (2009). Rather than merely using the average values of $k$ and $\sigma$ quoted by 
Brumley {\it et al}. ($k = 4.7$, $\sigma = 203~$rad/s), we use the individual values for each of the 60 \emph{Volvox} colonies from which the 
averages were obtained, together with their radii $a$. We also need the value of the dimensionless amplitude $\epsilon$. As discussed above, the 
recorded radius $a_0$ is the radius of the surface of the extra-cellular matrix in which the somatic cells are embedded, and $a = a_0 + 2L/3$ and 
hence that $\epsilon = L/(3a_0+2L)\approx L/3a_0$  (noting the typical orbit in figure~\ref{envelope}(b)). \cite{Solari:etal2011} have shown that 
flagellar length, as well as colony radius, increases as a colony of {\it V.carteri} or {\it V.barberi} ages. The values of $L$ (14.9 $\mu$m - 20.5 
$\mu$m) and $a_0$ quoted by them give values of $\epsilon$ between 0.029 and 0.038; thus we may be justified in choosing $\epsilon = 0.035$ 
as normal. We also use the value of $\delta$ (1.68) quoted above, although trajectories of flagellar tips measured by Brumley {\it et al}. (2014) 
show a range of values of $\delta$ from 1.45 to 1.86. Moreover we use $\tau = \tan(20^\circ) \approx 0.36$ although we do not have measurements 
of the offset angle for individual colonies. 

\begin{figure}
\begin{center}
\includegraphics[width=\textwidth]{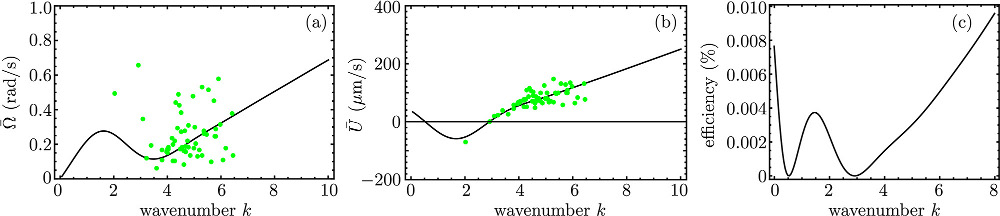}
\end{center}
\caption{Predicted values of (a) mean angular velocity $\bar{\Omega}$, (b) mean swimming speed $\bar{U}$ and (c) mechanical efficiency, $E$, 
as functions of the metachronal wavenumber $k$. Green dots are predictions of the squirmer model using the individually measured parameters for 
each of the 60 {\it Volvox} colonies. The solid lines are the predictions using the mean properties ($k=4.7$, $\sigma=203~$rad$/$s). Other 
parameters include $\delta=1.68$, $\chi=-\pi/2$, $\tau = \tan(20^\circ)$. Here the mean amplitude is $\epsilon \approx 0.05$, equivalent to 
flagella length $L=20~\mu$m.} \label{U_Omega_vs_k_L20}
\end{figure}

The results for $\bar{U}$ ($= \epsilon^2 \bar{U}^{(2)}$) and $\bar{\Omega}$ ($= \epsilon^2 \bar{\Omega}^{(2)}$) are plotted against $k$ in figure~\ref{U_Omega_vs_k_L20}, where the dots use the individual values of $k$, $\sigma$ and $a$ in each of the 60 {\it Volvox} colonies measured 
by Brumley {\it et al.} (2015). The continuous curve uses the mean values of $\sigma$ and $a$; all results assume a flagellum of length $L=20~\mu$m, 
i.e. a mean value of $\epsilon$ of 0.035. It is interesting that $\bar{U}$ and, to a lesser extent, $\bar{\Omega}$ increase regularly with $k$ over the 
range of measured values, but would vary considerably for lower values, even resulting in negative mean swimming speeds. 

Also plotted, in figure~\ref{U_Omega_vs_k_L20}(c), is the mechanical efficiency
\begin{equation}
E = 6 \pi \mu a \bar{U}^2 / \bar{P},
\end{equation}
where $P$ is the instantaneous rate of working of the stresses at the surface of the sphere,
\begin{equation}
P = 2 \pi a^2 \int_0^{\pi} \big( u_r \sigma_{rr} + u_{\theta}\sigma_{r \theta} + u_{\phi} \sigma_{r \phi} \big) \sin \theta_0 d \theta_0,
\end{equation}
and $\sigma$ is the stress tensor. The formula for $P$ in the absence of swirl was given by \cite{Blake:squirmer}, Eq.~(9); the additional, third, 
term due to swirl is equal to
\begin{equation}
16 \mu \pi a^3 \sum_{n=2}^{\infty} \frac{(n+2)}{n(n+1)(2n+1)} C_n^2 \label{eq:3.13}
\end{equation}
(see also \citet{PakLauga2014}).
Figure~\ref{U_Omega_vs_k_L20}(c) shows a local maximum of $E$ at $k \simeq 1.5$, corresponding to negative swimming speed, which may 
therefore be discounted. For $k>3.0$, however, the efficiency increases with $k$. According to this model, then, it appears that the swimming 
mode of {\it Volvox} did not come about through energetic optimisation. 

\begin{figure}
\begin{center}
\includegraphics[width=\textwidth]{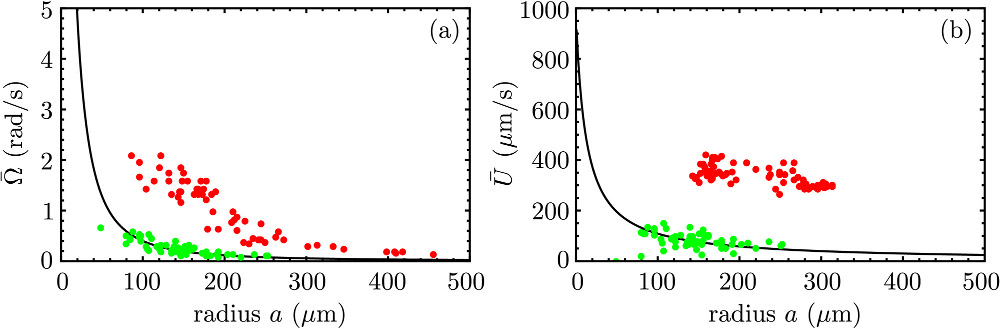}
\end{center}
\caption{Predicted and measured values of (a) mean angular velocity $\bar{\Omega}$ and (b) mean swimming speed $\bar{U}$, as functions of 
colony radius. Green dots are predictions of this model, red dots are measurements (on a different population of colonies) by \cite{Drescher:2009vn} 
({\it cf.} figure~\ref{Drescher_results}). Solid line is the prediction from mean properties of the 60 colonies whose metachronal wave data have been 
used.} \label{U_Omega_L20}
\end{figure}

We plot the calculated $\bar{U}$ and $\bar{\Omega}$ against $a$ in figure~\ref{U_Omega_L20}. The green points represent colony-specific 
predictions using data from Brumley {\it et al.} (2015) and the continuous curves correspond to the mean values of $k$, $\sigma$ and $\epsilon$ 
referred to above. The red points represent the experimental values measured by Drescher {\it et al}. (2009), again using the individual values of 
$\bar{U}$, $\bar{\Omega}$ and $a$ for each of the colonies measured (data kindly supplied by Dr. Knut Drescher) rather than an average value. 
As noted in the introduction, with reference to figure~\ref{Drescher_results}, because the above theory assumes neutral buoyancy, the value 
quoted for $U$ is the sum of the actual upwards swimming speed $U_1$ and the sedimentation speed $V$ of an inactive colony of the same radius. 

\begin{figure}
\begin{center}
\includegraphics[width=\textwidth]{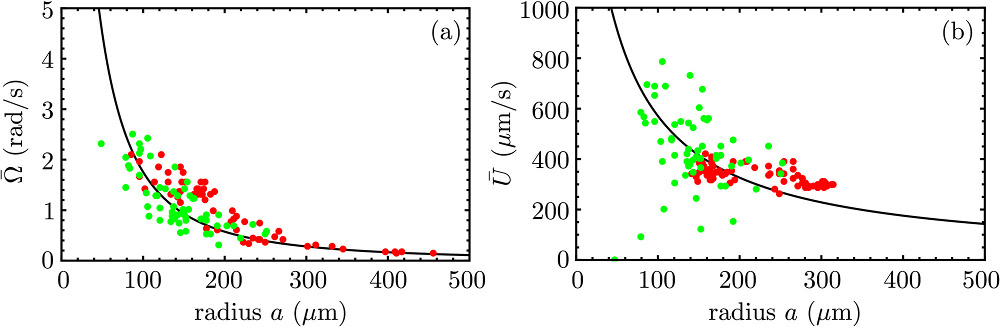}
\end{center}
\caption{Same as figure~\ref{U_Omega_L20} but with mean $\epsilon \approx 0.10$ ($L = 50~\mu$m).} \label{U_Omega_L50}
\end{figure}

In figure~\ref{U_Omega_L20}, the predictions for both $\bar{U}$ and $\bar{\Omega}$ are significantly below the measured values, though the 
trend with increasing radius is similar. If we had taken the flagellar length $L$ to be $50~\mu$m instead of $20~\mu$m, the agreement would 
seem to be almost perfect (figure~\ref{U_Omega_L50}). In the next section we discuss in more detail aspects of the model that may need to be 
improved.

\begin{figure}
\begin{center}
\includegraphics[width=0.75\textwidth]{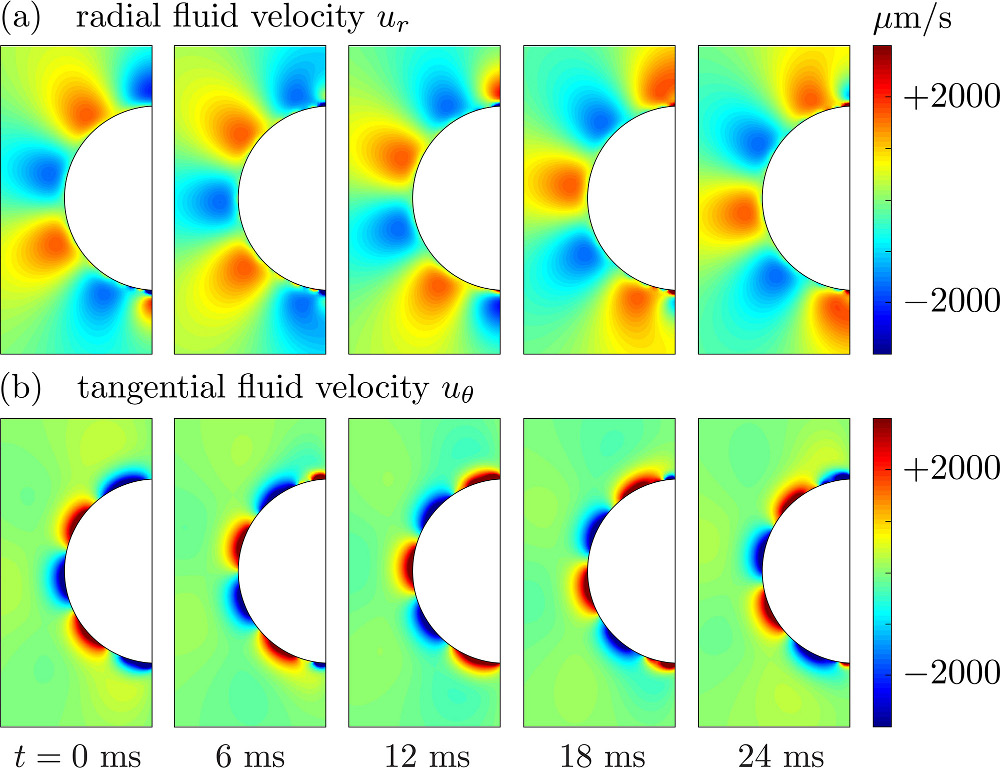}
\end{center}
\caption{Squirming flow fields. Radial (a) and tangential (b) components of the fluid velocity field shown at various times through one 
flagellar beating cycle. The metachronal wave properties (Eqs.~(\ref{eq:3.1}) and (\ref{eq:3.2})) are the same as for the average 
{\it Volvox} colony ($k=4.7$, $\sigma=203~$rad$/$s, $a_0=144~\mu$m) and other parameters correspond to measured flagella 
and their trajectories ($L=20~\mu$m, $\delta=1.68$, $\chi=-\pi/2$).} \label{squirmer_snapshots}
\end{figure}

\begin{figure}
\begin{center}
\includegraphics[width=0.8\textwidth]{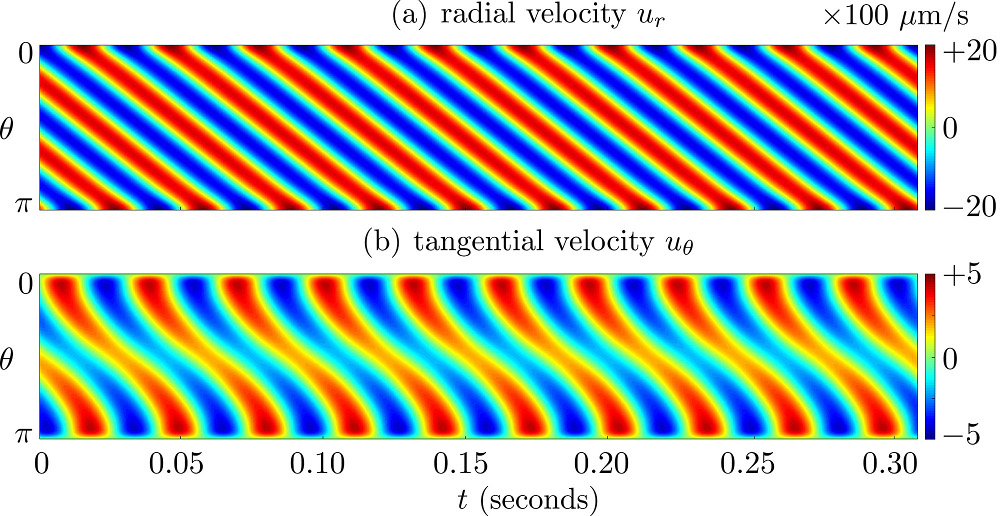}
\end{center}
\caption{Squirmer kymographs. Radial (a) and tangential (b) components of the flow, as functions of polar angle $\theta$ and time $t$, 
computed at the fixed radius ($r=1.3 \times a_0$). Other parameters are the same as in figure~\ref{squirmer_snapshots}.} 
\label{squirmer_kymographs}
\end{figure}

In addition to calculating $\bar{\Omega}$ and $\bar{U}$ we can use the squirmer model to compute the time-dependent velocity field, for 
comparison with the measurements in figures~\ref{exp_flow} and \ref{exp_kymographs}. Figure~\ref{squirmer_snapshots} shows the radial 
and tangential velocities as functions of position at different times during a cycle, for the mean values of $k$ (4.7), $\sigma$ (203 rad/s) and 
$a_0 = 144~\mu$m. Both velocity components show the metachronal wave, which is not surprising since that was used as input from 
Eqs.~(\ref{eq:3.1}) and (\ref{eq:3.2}). The figure also indicates that the tangential velocity component decays more rapidly with radial 
distance than the radial component. Calculated kymographs of $u_r$ and $u_{\theta}$ at $r = 1.3 \times a_0$ are shown in figure~\ref{squirmer_kymographs}, 
and can be compared with figure~\ref{exp_kymographs}. There is good qualitative agreement between figures~\ref{squirmer_snapshots} and \ref{squirmer_kymographs} and figures~\ref{exp_flow} and \ref{exp_kymographs}. Unlike the mean 
velocity, however, which is lower than measured, the amplitude of the calculated $u_r$ or $u_\theta$ oscillations, scaling as $\sigma a_0 \epsilon$ 
from Eqs.~(\ref{eq:2.11}) and (\ref{eq:3.1}), is about $1000$ $\mu$m s$^{-1}$, significantly larger than the measured value of about 
$300$ $\mu$m s$^{-1}$ (figure~\ref{exp_kymographs}). 

\section{Discussion}\label{sec:4}

The main discrepancy between the theoretical predictions of this paper and the experimental observations of Drescher {\it et al}. (2009) is that, 
although the maximum fluid velocity during a cycle, for the experimental parameter values, is much larger in the model than measured, the 
predicted mean velocity and angular velocity are significantly smaller than measured. 

The envelope model is clearly a great oversimplification, because even in the context of single-celled ciliates, the cilia tips do not form a continuous 
surface at all times. Not only may there be wide spaces between neighbouring tips, but also some tips may, during their recovery stroke, be 
overshadowed by others in their power stroke, so the envelope is not single-valued (\cite{Brennen:ARFM}). The latter is not a problem for 
\emph{Volvox}, because the flagellar pairs are more widely spaced, but that in itself adds to the former difficulty. \cite{Blake:squirmer} argued 
that the envelope model would be a better approximation for symplectic metachronal waves than for antiplectic ones, because the tips are closer 
together during the power stroke, when their effect on the neighbouring fluid is greatest; this is especially true for a ciliate such as \emph{Opalina}, 
but is less compelling in the case of \emph{Volvox}, for which typical cell (and hence flagellar) spacings are roughly equal to the flagellar length. The 
wide spacing between flagellar tips means that much of the `envelope' is not actively engaged in driving fluid past the surface, and fluid can leak back 
between neighbours, so one would expect the model to overestimate the fluid velocity, as it does if one considers the maximum instantaneous radial or tangential velocity.

Why, therefore, is the mean velocity underestimated? It seems likely that the difference lies in the fact that each flagellum beats close to the no-slip 
surface of the extracellular matrix in which the somatic cells are embedded. In the power stroke, a flagellum is extended and its outer parts, in particular 
the tip, set neighbouring fluid particles in motion, over a range of several flagellar radii, at about the same speed as the tip. During the recovery stroke, 
on the other hand, the flagellum is much more curved, and the outer part remains roughly parallel to the colony surface. Thus the drag exerted by this 
part of the flagellum on the fluid will be reduced by a factor approaching 2 compared with the power stroke. Moreover, this outer part is relatively close 
to the colony surface, and the no-slip condition on that surface will prevent fluid particles from moving at the same speed as the tip except very close 
to it. Both these factors mean that, although every element of the beating flagellum oscillates with zero mean, the fluid velocities that it generates do not. 

\begin{figure}
\begin{center}
\includegraphics[width=0.9\textwidth]{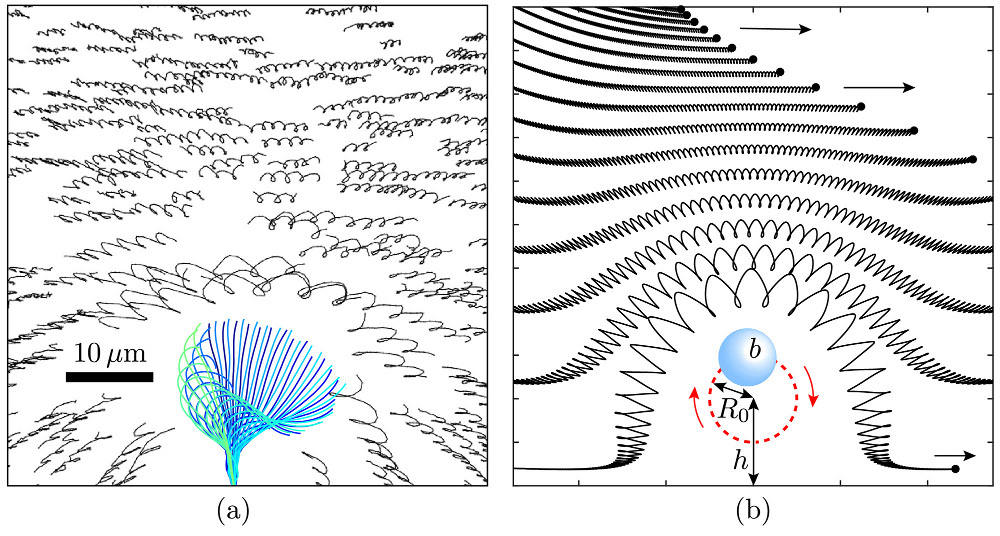}
\end{center}
\caption{Particle paths in the vicinity of a flagellum. (a) Trajectories of $0.5\,\mu$m passive tracers near an isolated {\it Volvox} flagellum held with 
a glass micropipette. The tracked flagellar waveform from several beats is also shown. (b) A sphere of radius $b$ moving in a circular trajectory 
above and perpendicular to a no-slip boundary produces a time-dependent flow, which closely mimics that of a real flagellum. This simulation of 
100 beats shows particle paths from various initial positions, and corresponds to $h=10\, \mu$m, $R_0 = 5\, \mu$m.} 
\label{particle_trajectories}
\end{figure}

As part of the experiments reported by \cite{Brumley2014}, movies were taken of the motion of microspheres in the flow driven by a single beating 
flagellum on an isolated \emph{Volvox} somatic cell fixed on a micropipette. Experimental details are given briefly in appendix~\ref{appC}. One of 
these movies is reproduced in supp. mat. movie S3, in which the difference between the fluid particle displacements in power and recovery strokes 
can be clearly seen. The trajectories of a number of the microspheres are shown in figure~\ref{particle_trajectories}(a). Supp. mat. movie S4 and figure~\ref{particle_trajectories}(b) show particle trajectories calculated from a very simple model (see appendix~\ref{appC}), which consists of a 
small spherical bead following a circular orbit perpendicular to a nearby rigid plane (such an orbiting bead model of a beating flagellum has been used extensively in recent years; \cite{Lenz:2006fk,Vilfan:2006uq,Niedermayer:2008fk,Uchida:2011kx,Brumley:2012,Brumley:2015MW,Bruot2015}). The 
similarity between the measured and computed trajectories is clear.

It is therefore evident that the net tangential velocity excess of the power stroke over the recovery stroke of \emph{Volvox} flagella will be 
$O(\epsilon)$, so the mean velocity generated will be $O(\epsilon)$ not $O(\epsilon^2)$ as obtained from our squirmer model. That may be 
a more important limitation of the model than the wide spacing of the flagella. What is required, in future, is a detailed fluid dynamic analysis 
of an array of beating flagella on the surface of a sphere. This will be an extension of the so-called sublayer model of \cite{Blake:1972ciliamodel} 
and \cite{Brennen:ARFM}, in which each cilium is represented as a linear distribution of Stokeslets whose strengths can be estimated using resistive 
force theory, or calculated more accurately as the solution of an integral equation using slender-body theory, taking account of the no-slip boundary 
by including the Stokeslet image system as derived for a planar boundary by \cite{Blake:image}. Such a model of an array of cilia on a planar 
boundary has been studied by \citet{Dingetal} and used to study fluid mixing and solute transport. The generalization to a sphere is currently being developed.

An additional assumption of the theory of this paper is that the elliptical trajectory of each flagellar tip has its major axis parallel to the locally 
planar no-slip colony surface. In fact it will in general be at a non-zero angle $\psi$ to that surface (figure~\ref{envelope}(b)). In that case the 
calculation becomes somewhat more cumbersome but no more difficult, as outlined in appendix~\ref{appD}. If we choose $\psi = 30^{\circ}$, 
for example, the results for $\bar{U}$ and $\bar{\Omega}$ are negligibly different from those in figure~\ref{U_Omega_L20}. The assumption 
that $\psi = 0$ is therefore not responsible for the discrepancy between theory and experiment in that figure.

\section*{Acknowledgements}
The authors are very grateful to Dr Knut Drescher, for the use of his original data in figure~8, Dr Kirsty Wan, for her data in figure~6(b), and 
Dr Thomas Montenegro-Johnson, for enlightening discussions on the future development of a complete sublayer model of \emph{Volvox} swimming. 
This work was supported by a Human Frontier Science Program Cross-Disciplinary Fellowship (D.R.B.) and a Senior Investigator Award from the 
Wellcome Trust (R.E.G.).

\appendix
\section{Integrals required in the derivation of Eq.~(\ref{eq:2.16})}\label{appA}

We seek to evaluate
\begin{equation}\label{eq:A.1}
J_{nm} = \int_{0}^{\pi}\sin^2{\theta_0} P_n(\cos{\theta_0}) V_m(\cos{\theta_0})~d\theta_0
\end{equation}
and
\begin{equation}\label{eq:A.2}
K_{nm} = \int_{0}^{\pi}\sin{\theta_0}\cos{\theta_0} V_n(\cos{\theta_0}) V_m(\cos{\theta_0})~d\theta_0,
\end{equation}
where $V_n$ is defined by (\ref{eq:2.2}), using the standard recurrence relations and differential equation for Legendre polynomials:
\begin{equation}\label{eq:A.3}
xP_{n}' = nP_{n} + P_{n-1}'
\end{equation}
\begin{equation}\label{eq:A.4}
(2n+1)xP_{n} = (n+1)P_{n+1} + nP_{n-1}
\end{equation}
\begin{equation}\label{eq:A.5}
\frac{d}{dx}\left[(1-x^2)P_{n}'\right] = -n(n+1)P_{n}.
\end{equation}
Here a prime means $\frac{d}{dx}$ and we do not explicitly give the $x$-dependence of $P_{n}(x)$. From (\ref{eq:A.1}),
\begin{equation}\label{eq:A.6}
J_{nm} = \frac{2}{m(m+1)}\int_{-1}^{1}P_{n}(1-x^2)P_{m}'~dx = 2\int_{-1}^{1}I_{n}(x)P_{n}~dx ~~  \mathrm{(by~ parts)}
\end{equation}
where
\begin{equation}\label{eq:A.7}
I_{n}(x) = \int^{x} P_{n}~dx = \frac{xP_{n} - P_{n-1}}{n+1}.
\end{equation}
Hence
\begin{equation}\label{eq:A.8}
J_{nm} = \frac{2}{2n+1}\int_{-1}^{1}P_{m}(P_{n+1} - P_{n-1})~dx = \frac{4}{2n+1}\left(\frac{\delta_{m,n+1}}{2n+3} - \frac{\delta_{m,n-1}}{2n-1}\right).
\end{equation}
From (\ref{eq:A.2}),
\begin{eqnarray}\label{eq:A.9}
K_{nm} &=& \frac{4}{n(n+1)m(m+1)}\int_{-1}^{1}xP_{n}'(1-x^2)P_{m}'~dx \nonumber\\
       &=& \frac{4}{n(n+1)}\int_{-1}^{1}(nI_{n} + P_{n-1})P_{m}~dx  ~~\mathrm{(by~ parts~ and~ using~ (\ref{eq:A.3}))} \nonumber\\ &=&  
\frac{4}{n(n+1)}\int_{-1}^{1}\left(\frac{n}{2n+1}P_{n+1} + \frac{n+1}{2n+1}P_{n-1}\right)P_{m}~dx ~~ 
\mathrm{(using ~(\ref{eq:A.4}))} \nonumber\\ &=& \frac{8}{2n+1}\left[\frac{\delta_{m,n+1}}{(n+1)(2n+3)} + \frac{\delta_{m,n-1}}{n(2n-1)}\right].
\end{eqnarray}

\section{Proof of Eq.~(\ref{eq:3.5a})}\label{appB}
We prove by induction the first of the formulae in Eq.~(\ref{eq:3.5}); proofs of the others are similar.
Let
\begin{equation}\label{eq:B.1}
Q_{n}(k) = \int_{0}^{\pi}\sin{\theta}P_{n}(\cos{\theta})\sin{k\theta}~d\theta,
\end{equation}
so that
\begin{equation}\label{eq:B.2}
a_{n}^{(s)} = \frac{2n+1}{2}Q_{n}(k),
\end{equation}
from the first of (\ref{eq:3.4a}). The result we seek to prove is
\begin{equation}\label{eq:B.3}
Q_{n}(k) = (-1)^{n} 2k\left[(-1)^{n+1} + ~\cos{k\pi}\right]\eta(k,n),
\end{equation}
where $\eta(k,n)$ is given by (\ref{eq:3.6}). From (\ref{eq:B.1}) and (\ref{eq:A.4}), we have
\begin{eqnarray}\label{eq:B.4}
Q_{n+1}(k) &=& \int_{0}^{\pi} ~\sin{k\theta}~\sin{\theta} \left[\frac{2n+1}{n+1}\cos{\theta}~P_{n} - \frac{n}{n+1}P_{n-1}\right]~d\theta \nonumber\\ &=& -\frac{n}{n+1}Q_{n-1}(k) + \frac{2n+1}{n+1}\int_{0}^{\pi} ~\sin{k\theta}~\sin{\theta}~\cos{\theta}~P_{n}~d\theta \nonumber\\ &=& -\frac{n}{n+1}Q_{n-1}(k) + \frac{2n+1}{2(n+1)}\int_{0}^{\pi} \left[\sin{(k+1)\theta} + \sin{(k-1)\theta}\right]\sin{\theta}~P_{n}~d\theta \nonumber\\ &=& -\frac{n}{n+1}Q_{n-1}(k) + \frac{2n+1}{2(n+1)}\left[Q_{n}(k+1) + Q_{n}(k-1)\right].
\end{eqnarray}
Now suppose that (\ref{eq:B.3}) is true for $Q_{n-1}$ and $Q_n$, for all $k$, substitute it into the right hand side of (\ref{eq:B.4}), and after some algebra indeed obtain (\ref{eq:B.3}) with $n$ replaced by $n+1$. The induction can be shown to start, with $n=1$ and $n=2$, using the standard identities
\begin{equation}\label{eq:B.5}
\Gamma(z+1) = z\Gamma(z)
\end{equation}
\begin{equation}\label{eq:B.6}
\Gamma(z)\Gamma(1-z) = -z\Gamma(-z)\Gamma(z) = \frac{\pi}{~\sin{(\pi z)}}.
\end{equation}
Thus (\ref{eq:B.3}) and hence (\ref{eq:3.5a}) are proved.

\section{Flagellar flow fields}\label{appC}

To investigate the time-dependent flow fields produced by individual eukaryotic flagella, \cite{Brumley2014} isolated individual cells from colonies 
of {\it Volvox carteri}, captured and oriented them using glass micropipettes, and imaged the motion of $0.5\, \mu$m polystyrene microspheres 
within the fluid at 1000~fps. One such movie is included as supp. mat. movie S3, which shows the time-dependent motion of these passive tracers 
in the vicinity of the beating flagellum. We identify the trajectories of the microspheres, and these are shown in figure~\ref{particle_trajectories}(a), together with the tracked flagellar waveform over several beats. Tracer particles in the immediate vicinity of the flagellar tip exhibit very little back flow during the recovery stroke.

We consider now the flow field produced by a simple model flagellum, which consists of a sphere of radius $b$ driven at a constant angular speed 
$\omega$ around a circular trajectory of radius $R_0$, perpendicular to an infinite no-slip boundary. The trajectory of the sphere is given by 
\begin{equation}
\mathbf{x}_1(t) = \mathbf{x}_0 + R_0 \big( \cos \omega t \, \hat{\mathbf{z}} + \sin \omega t  \, \hat{\mathbf{y}} \big)
\end{equation}
where $\mathbf{x}_0 = h \, \hat{\mathbf{z}}$. The velocity of the particle is then
\begin{equation}
\mathbf{v}_1 = \dot{\mathbf{x}}_1 = \omega R_0 \big( -\sin \omega t \, \hat{\mathbf{z}} + \cos \omega t  \, \hat{\mathbf{y}} \big).
\end{equation}
The force that this particle imparts on the fluid is given by
\begin{equation}
\mathbf{F}_1 = \mathbf{\gamma}_1 \cdot \mathbf{v}_1 = \gamma_0 \bigg[\mathbf{I} + \frac{9b}{16z(t)} (\mathbf{I} + 
\hat{\mathbf{z}}\hat{\mathbf{z}}) \bigg] \cdot \mathbf{v}_1.
\end{equation}
We know that $z(t) = h + R_0 \cos \omega t$, and therefore the time-dependent force exerted on the fluid is
\begin{equation}
\mathbf{F}_1(t) = \gamma_0 \omega R_0 \bigg[ \cos \omega t \, \hat{\mathbf{y}} - \sin \omega t \, \hat{\mathbf{z}} + \frac{9b}{16 
(h + R_0 \cos \omega t)} \big( \cos \omega t \, \hat{\mathbf{y}} - 2 \sin \omega t \, \hat{\mathbf{z}} \big) \bigg]. \label{app_F1t} 
\end{equation}
The fluid velocity $\mathbf{u}(\mathbf{x})$ at position $\mathbf{x}$ is expressed in terms of the Green's function in the presence of the 
no-slip boundary condition (\cite{Blake:image}):
\begin{equation}
\mathbf{u}(\mathbf{x}) = \mathbf{G}(\mathbf{x}_1(t), \mathbf{x}) \cdot \mathbf{F}_1(t)
\end{equation}
where 
\begin{equation}
\textbf{G}(\mathbf{x}_i,\mathbf{x}) = \textbf{G}^\textit{S}(\mathbf{x} - \mathbf{x}_i) - \textbf{G}^\textit{S}(\mathbf{x} - \bar{\mathbf{x}}_i) + 2 z_i^2 \textbf{G}^\textit{D}(\mathbf{x} - \bar{\mathbf{x}}_i) - 2 z_i \textbf{G}^\textit{SD}(\mathbf{x} - \bar{\mathbf{x}}_i)
\end{equation}
and
\begin{eqnarray}
G_{\alpha \beta}^\textit{S}(\mathbf{x}) &&= \frac{1}{8 \pi \mu} \bigg(\frac{\delta_{\alpha \beta}}{|\mathbf{x}|} + \frac{\mathbf{x}_{\alpha} \mathbf{x}_{\beta}}{|\mathbf{x}|^3} \bigg), \label{app_GS} \\
G_{\alpha \beta}^\textit{D}(\mathbf{x}) &&= \frac{1}{8 \pi \mu} \big( 1 - 2 \delta_{\beta z} \big) \frac{\partial}{\partial x_{\beta}} \bigg( \frac{x_{\alpha}}{|\mathbf{x}|^3} \bigg), \label{app_GD} \\
G_{\alpha \beta}^\textit{SD}(\mathbf{x}) &&= \big( 1 - 2 \delta_{\beta z} \big) \frac{\partial}{\partial x_{\beta}} G_{\alpha z}^S (\mathbf{x}). \label{app_GSD} 
\end{eqnarray}
For a passive tracer with initial position $\mathbf{x}=\mathbf{X}_0$ at $t=t_0$, its trajectory can be calculated according to
\begin{equation}
\mathbf{x}(t) - \mathbf{X}_0 = \int_{t_0}^{t} \mathbf{G}\big(\mathbf{x}_1(\tau), \mathbf{x}(\tau)\big) \cdot \mathbf{F}_1(\tau) \, d \tau. \label{app_particle_path}
\end{equation}
Numerical solutions of Eq.~(\ref{app_particle_path}) are shown in figure~\ref{particle_trajectories}(b) for various initial positions. The 
parameters used are designed to mimic those of real {\it Volvox} flagella ($h=10 \, \mu$m, $R_0 = 5 \, \mu$m). A sphere of radius 
$b=5 \, \mu$m is used, though we emphasise that strictly speaking this does not come into contact with the plane. The finite value of $b$ 
is used simply to generate variable drag as a function of height, in order to produce a net flow. Additionally, the particle trajectories are independent 
of the speed of the sphere, and so the results in figure~\ref{particle_trajectories}(b) would be unchanged if the sphere were instead driven by either a constant force, or by a phase-dependent term. 

\section{Rotated ellipse}\label{appD}

In this section, we consider the case in which the elliptical trajectory of the flagellar tip is rotated at an angle $\psi$ with respect to the surface of the {\it Volvox} colony. In this case, Eqs.~(\ref{eq:3.1}) and (\ref{eq:3.2}) can be generalised to become 
\begin{eqnarray}
&& R-a =  \cos \psi \big[ a \epsilon \sin (k \theta_0-\sigma t) \big] -  \sin \psi \big[ a \epsilon \delta \sin (k \theta_0-\sigma t - \chi) \big], \\
&& \theta-\theta_0 = \cos \psi \big[ \epsilon \delta \sin (k \theta_0-\sigma t-\chi) \big] +  \sin \psi \big[ \epsilon \sin (k \theta_0-\sigma t) \big].
\end{eqnarray}
The series expansions for these are then given by
\begin{eqnarray}
&& \sum_{n=0}^{\infty} \alpha_n(t) P_n(\cos \theta_0) = \cos \psi \sin(k \theta_0 - \sigma t) - \delta  \sin \psi \sin(k \theta_0 - \sigma t - \chi), \\
&& \sum_{n=1}^{\infty} \beta_n(t) V_n(\cos \theta_0) = \delta \cos \psi \sin(k \theta_0 - \sigma t - \chi) + \sin \psi \sin(k \theta_0 - \sigma t),
\end{eqnarray}
and $\gamma_n(t) = \tau \beta_n(t)$ as before. Equations~(\ref{eq:A.3}) and (\ref{eq:A.4}) need to be solved for $\alpha_n$ and $\beta_n$, but 
this follows easily by linearity using the solutions in Eqs.~(\ref{eq:3.7a}) and (\ref{eq:3.7b}), together with appropriate transformations in $t$. 
Calculation of $\bar{\Omega}^{(2)}$ and $\bar{U}^{(2)}$ is more challenging, but after considerable algebra, we find the following:
\begin{eqnarray}
&& \bar{\Omega}^{(2)}  =  18 \sigma  \tau \eta(k,1)\eta(k,2) \sin k \pi [ (\delta ^2-1)  \cos 2\psi + 1+ \delta ^2 +2 \delta  \cos \chi \sin 2\psi] \nonumber \\
&& + \frac{3}{2}\sigma  \tau  \sin k\pi \sum_{n=2}^{\infty} \eta(k,n)\eta(k,n+1) (-1)^{n+1}(n+1)(n+2) \bigg[ n(n+1)(\delta ^2-1) \cos 2\psi \nonumber \\
&& \qquad \qquad + k(2n+3)\delta \sin \chi + n(n+1) (1+\delta ^2 + 2\delta  \cos \chi \sin 2\psi) \bigg], \label{Omega_rotated_appendix}
\end{eqnarray}
and
\begin{eqnarray}
&& \bar{U}^{(2)} = - 6 a \sigma \eta(k,1)\eta(k,2) \sin k\pi \bigg[ \frac{3 \delta  \sin \chi}{k}  +2 (\delta ^2 + 2 \delta  \cos \chi \sin 2\psi +1)+2(\delta ^2-1) \cos 2\psi \bigg] \nonumber \\
&& + \frac{1}{2} a \sigma  \sin k\pi \sum_{n=2}^{\infty} (-1)^n \eta(k,n)\eta(k,n+1) \bigg[ 4 k \delta (n+1) (2n^2+3n+2) \sin \chi \nonumber \\
&&  \qquad \qquad + k^2 (2n^2-2n-1) \big[ (\delta ^2-1) \cos 2\psi -\delta ^2 + 2 \delta  \cos\chi \sin 2\psi -1 \big]  \nonumber \\
&&   \qquad \qquad +2 n (n+2)(n+1)^2 \big[ (\delta ^2-1) \cos 2\psi  +\delta ^2 + 2 \delta  \cos\chi \sin 2\psi +1) \big] \bigg]. \label{U_rotated_appendix}
\end{eqnarray}
Note that Eqs.~(\ref{Omega_rotated_appendix}) and (\ref{U_rotated_appendix}) reduce to Eqs.~(\ref{eq:3.8}) and (\ref{eq:3.9}) respectively when $\psi = 0$.

\newpage
\bibliographystyle{jfm}
\textit{}\bibliography{tjpbibl}

\newpage

\noindent{\large\bf Supplementary Movies}

\medskip

\noindent{\bf S1:} Radial component of fluid velocity field, measured with PIV, as a function of
polar angle $\theta$ and time (cf. figure 4b).

\smallskip

\noindent{\bf S2:} Tangential component of fluid velocity field, measured with PIV, as a function
of polar angle $\theta$ and time (cf. figure 4c).

\smallskip

\noindent{\bf S3:} Observed particle motions in the vicinity of a single beating flagellum on an
isolated \emph{Volvox} somatic cell (cf. figure 12a).

\smallskip

\noindent{\bf S4:} Computed particle trajectories generated by a microsphere moving in a circular
orbit above and perpendicular to a no-slip boundary (cf. figure 12b).

\end{document}